\documentclass{jpp}
\usepackage{graphicx}
\usepackage{color}
\usepackage[utf8]{inputenc}
\usepackage[T1]{fontenc}
\usepackage{amsmath}
\usepackage{color}
\usepackage{soul,xcolor}
\usepackage[normalem]{ulem}

\shorttitle{Non-Maxwellian rate coefficients for electron and ion collisions}
\shortauthor{D. Vrinceanu, R. Onofrio and H.R. Sadeghpour}

\title{Non-Maxwellian rate coefficients for electron and ion collisions in Rydberg plasmas: Implications for excitation and ionization}

\author{Daniel Vrinceanu\aff{1},
  Roberto Onofrio\aff{2,3}
  \corresp{\email{onofrior@gmail.com}}
 \and H. R. Sadeghpour\aff{4}}

\affiliation{\aff{1}
Department of Physics, Texas Southern University, Houston, TX 77004, USA
\aff{2}Dipartimento di Fisica e Astronomia ``Galileo Galilei'', \\ Universit\`a di Padova, 
Via Marzolo 8, 35131 Padova, Italy
\aff{3}Department of Physics and Astronomy, 6127 Wilder Laboratory, \\ Dartmouth College, Hanover 03755, USA
\aff{4}ITAMP, Harvard-Smithsonian Center for Astrophysics, Cambridge, MA 02138, USA}
\begin{document}

\maketitle

\begin{abstract}
Scattering phenomena between charged particles and highly excited Rydberg atoms are of critical importance in many processes in plasma physics and astrophysics. While a Maxwell-Boltzmann (MB) energy distribution for the charged particles is often assumed for calculations of collisional rate coefficients, in this contribution we relax this assumption and use two different energy distributions, a bimodal MB distribution and a $\kappa$-distribution. Both variants share a high-energy tails occurring with higher probability than the corresponding MB distribution. The high energy tail may significantly affect rate coefficients for various processes. We focus the analysis to specific situations by showing the dependence of the rate coefficients on the principal quantum number of hydrogen atoms in $n$-changing collisions with electrons in the excitation and ionization channels and in a temperature range relevant to the divertor region of a tokamak device. We finally discuss the implications for diagnostics of laboratory plasmas.
\end{abstract}

\section{Introduction}
Collisions of electrons and ions with neutral atoms are relevant in studies of 
stellar atmospheres \citep{mashonkina}, radio emission in recombination processes of H-II clouds, 
primordial cosmological recombination of hydrogen \citep{chluba}, and plasma fusion \citep{janev1,janev2}. 
Initial studies have focused on scattering and excitation of ground state or low-lying states, 
in particular for hydrogen atoms. More challenging is the extension to high-lying Rydberg states, for which {\it ab initio} quantum calculations become prohibitively untenable. This results in the use of various approximation schemes with unavoidable systematic errors which at times can be, in some observable quantities, of order 100 $\%$ \citep{rolfes,nagesha,przybilla,VOS2}. 

Among the various assumptions in these models, the fact that colliding particles may not share energies according to a MB probability distribution, as far as we know, has never been systematically scrutinized in the context of plasma fusion, apart from the possible impact in terms of nuclear fusion reactivities  \citep{onofrio}. This is at variance with the astrophysical plasmas for which deviations from MB have been discussed \citep{nicholls1,nicholls2,storey1,storey2,draine}. Deviations from the MB distributions are expected in a tokamak-confined plasma both in the Scrape-Off Layer (SOL), and in the divertor region, due to the lower densities and temperatures experienced by electrons and ions with respect to the core confinement region. The lowest density in these regions  ($10^{13}-10^{14}$ cm${}^{-3}$) implies that many-body dynamics closer to the collisionless regime, and therefore there can be deviations from MB due to lack of thermal equilibration. Moreover, Edge Localized Modes (ELMs) in the SOL region can suddenly release suprathermal particles spoiling a pre-existing MB distribution. 

In this work, we make a quantitative analysis of the electron-Rydberg hydrogen $n$-changing excitation and ionization processes with non-MB distributions, with $n$ the principal quantum number. The rate coefficient for generic collisional processes can be written as

\begin{equation}
k_{if} = \langle v \sigma_{if} \rangle = \int f(v)\; v\: \sigma_{if}(v) \; dv = 
\sqrt{\frac{2}{m}}
\int P(E) E^{1/2} \sigma_{if}(E)\; dE,
\label{rates}
\end{equation}
where $\sigma$ is the cross section for the given process, $f(v)$ is the probability distribution of the velocities of the projectile particles, and $P(E)$ is the corresponding probability energy distribution. The goal is to evaluate changes in the rate coefficients when distributions $f(v)$ and $P(E)$ in (\ref{rates}) deviate from MB.

Deviations from a MB distribution are investigated within two classes, a mixture of two MB distributions at different temperatures, and  the so-called $\kappa$-distribution. These two examples of non-Boltzmann distributions have in common  qualitative features, such as the presence of a substantial high-energy tail, and it is therefore interesting to study their quantitative impact with respect to a MB distribution for instance in collisions where the particles share the same low-energy distribution, or total internal energy. 

The range of energies and densities is chosen in order to characterize plasmas around the scrape-off layer of a tokamak machine and the divertor region, with electron densities, $\rho_e\sim 10^{13}$ cm${}^{-3}$ and energies, $E\sim 0.5-20$ eV \citep{anderson}. We describe in detail the two non-MB distributions in sections 2, and then proceed to discuss the results for excitation and ionization of hydrogen atoms of high $n$ in sections 3 and 4, respectively. In the concluding section, we relate the results to the atomic physics in the tokamak divertor region. 

\begin{figure}
\centering
\includegraphics[width=5.2in]{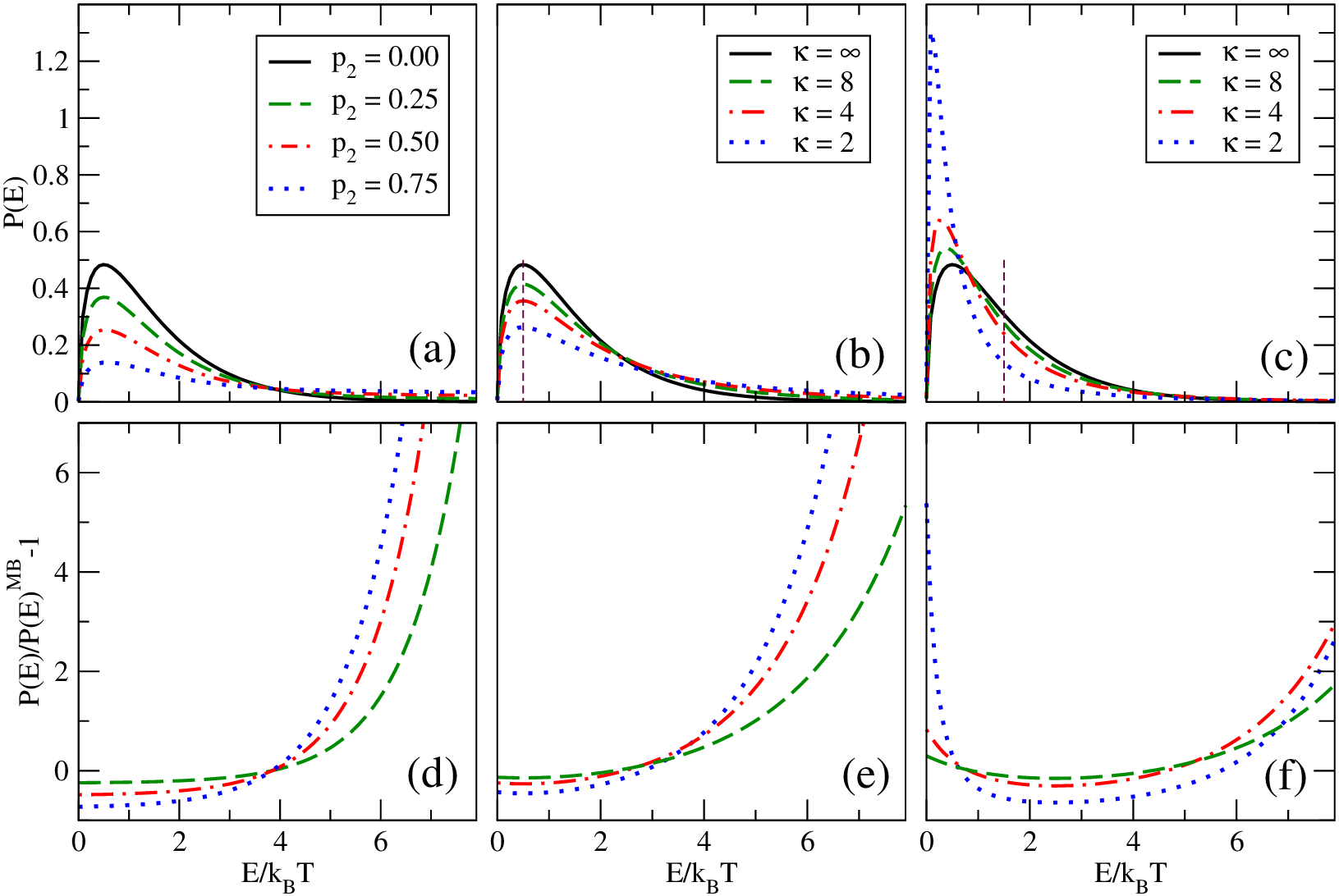}
\caption{Absolute (top panels) and relative (bottom panels) comparison of bMB and $\kappa$-distributions to MB distributions. The solid lines in the top panels represent the limiting cases for the MB distributions, i. e.  $p_2=0$, and $\kappa\rightarrow \infty$ for  the $\kappa$-distributions. (a) bMB distributions with increasing weight $p_2$ of the high-temperature component, and $T_2 = 10~ T_1$. (b) $\kappa$-distributions for $\eta = 1/2$ have maxima at $\langle E\rangle = {1\over2}k_B T$, indicated by  vertical dashed lines. (c) $\kappa$-distributions for $\eta = -3/2$ have average energy given by $\langle E\rangle = {3\over2} k_B T$, indicated by  vertical dashed lines. The bottom panels (d-f) show more clearly the relative deviations from a MB energy distribution for the corresponding top panels, especially in the high-energy tails. Notice that the $\kappa$-distributions for $\eta=1/2$ resemble more closely the bMB distributions, compared to the corresponding $\kappa$-distributions for $\eta=-3/2$. In the latter, the deviations from a MB distribution at high energy are less pronounced, while  significant deviations instead occur at low energy. The case of $\eta=0$, not depicted for graphical reasons, has a behavior close to the $\eta=1/2$ case.}
\label{Fig1}
\end{figure}

\section{Non-Maxwellian energy distributions}

We have chosen energy distributions differing from MB having in mind examples already available in various physical contexts which possess pronounced, ``hard'', high-energy tails. 
This feature can sensibly change the rate coefficients for processes with energy dependent cross-sections. In general, we expect formation of distributions with high-energy tail whenever there is an energy inflow into the system which is large enough to not be dissipated into all the modes during the typical relaxation time scales in the system \citep{livadiotis1,livadiotis2}.

The first example is provided by a mixture of two Maxwell-Boltzmann distributions (bMB) with different inverse temperatures $\beta_1=1/(k_BT_1)$ and $\beta_2=1/(k_BT_2)$

\begin{equation}
P_{bMB}(E) = 2\sqrt{\frac E\pi}\left[ p_1 \beta_1^{3/2} \exp(-\beta_1 E) + p_2 \beta_2^{3/2} \exp(-\beta_2 E) \right],
\label{bMB}
\end{equation}
where $p_1$ and $p_2$ are the statistical weights of the two distributions, {\it i.e.} the probability that a given particle will belong to 
the distribution with  $\beta_1$ or $\beta_2$ (where we assume, for instance, $\beta_1 > \beta_2$), respectively. Since $p_1+p_2=1$, the  probability distribution is characterized by three independent parameters, $\beta_1, \beta_2$ and either one of $p_1$ or $p_2$.
Having in mind cases in which the high-temperature component is not dominant, we choose $p_2$ such that in the $p_2=0$ case,  the MB distribution is recovered at $\beta_1$.

Such distributions appear in fusion plasma for instance after ion cyclotron resonance heating, as discussed in \citet{bhatnagar}. 
The neutral hydrogen flux energy distribution was measured, during heating, to contain a high-energy MB tail with a temperature of 48 keV on top of the pre-existing MB distribution at a temperature of 3 keV. Another example is provided by temperature anisotropy driven instabilities in the solar wind, for which the velocity distribution of the involved particles is adequately fitted by bMB distributions \citep{Klein1,Yoon,Klein2}. Shocks and winds affect the electron velocity distribution in collisionally-ionized plasmas, producing low-energy electrons with power-law tails \citep{Hahn}.
The rate coefficients corresponding to these bMB distributions are linear combinations of MB rate coefficients for temperatures $T_1$ and $T_2$, with weights given by $p_1$ and $p_2$.

More intriguing is the case of the $\kappa$-distributions. 
These are generalizations of Lorentzian distributions which were first introduced for applications in space plasma physics \citep{vasyliunas}. The characterization of these $\kappa$-distributions requires two parameters 
determining the shape and the corresponding temperature that can be associated with a probability distribution in a sense that is explained below.

The energy probability density is defined by
\begin{equation}\label{kappa}
P_{\kappa,\eta}(E) = 2\sqrt{\frac E\pi}\beta^{3/2} \frac{C(\kappa, \eta)}
{\left( 1 + \frac{\beta E}{\kappa + \eta}\right)^{\kappa + 1}}
\end{equation}
where $C(\kappa, \eta) = \Gamma(\kappa+1)/[\Gamma(\kappa-1/2)(\kappa + \eta)^{3/2}]$, $\kappa > 3/2$ and $\eta > -\kappa$. The MB distribution is recovered as $\kappa \rightarrow \infty$. According to a semiqualitative analysis presented in \citet{livadiotis3}, see in particular figure 3, 
any distribution with $\kappa > 20$ can be well approximated with an equilibrium MB distribution, while typical values for genuine $\kappa$-distributions out of thermal equilibrium require $\kappa$ in the range between about 2 and 10. Our choice of $\kappa$-parameters in the following considerations is based on this criterion.

Notice that we have introduced a parameter $\eta$ which allows to treat various $\kappa$-distributions differing in their average energy content. Indeed, the $\kappa$-distribution has a maximum at energy $E_{max} = (\kappa + \eta)/[(1+2\kappa)\beta]$ and an average energy $\langle E \rangle = 3(\kappa + \eta)/[(2\kappa - 3)\beta]$. This allows for various interpretations of the temperature associated with a $\kappa$-distribution.
For example, by setting $\eta = 1/2$, the distribution peaks at $E_{max} = 1/(2\beta) = k_B T/2$ independently of $\kappa$ and therefore in practical applications its temperature can be derived as $T = 2 E_{max}/k_B$. For $\eta = -3/2$, the average energy $\langle E \rangle$ is independent of $\kappa$, and temperature can be obtained as $T = 2 \langle E \rangle/ 3k_B$. Another interesting case is $\eta = 0$ for which the $\kappa$-distribution of velocities has a $\kappa$-independent maximum corresponding to the most probable velocity $v_{max} = \sqrt{2 k_B T/m}$ (see Appendix A for details). While from an operative standpoint, the values of $\eta$ and $\beta$ can be extracted by looking at the behaviour around the peak of the distribution, the value of $\kappa$ can be obtained by fitting the high-energy tail.

Figure \ref{Fig1} illustrates the main features of bMB and $\kappa$-distributions. They all have only one maximum and exhibit strong tails at large energies, particularly manifest in the lower plots, showing the relative difference from the corresponding MB distribution. For $\eta = 1/2$, the maximum of the distribution is at the same position, showing a significant fraction of particles with higher energy, as compared to MB distributions. At the other extreme, for $\eta = -3/2$, the average energy is the same for various values of $\kappa$, pushing a large number of particles towards energies lower than what would be expected for a MB distribution, thereby counterbalancing the still significant population in the high energy tail.

\section{Electron-Rydberg atom excitation rate coefficients}

Rydberg atom excitations due to electron scattering have been studied with different techniques and in various physical contexts, ranging from cold atomic plasmas in the laboratory \citep{rolfes,nagesha} to the primordial cosmological recombination \citep{chluba}. 
Since the pioneering experiment of Frank and Hertz, excitation by electron bombardment has been studied extensively for low-lying atomic states. However, there has been much more limited 
success, in terms of overall accuracy, in the case of Rydberg atomic states. 
This is not surprising because there is a vast gap between the case of low $n$, for which exact quantum mechanical computations are still feasible, and the case of high $n$ for which, based on the correspondence principle, Classical Trajectory Monte Carlo (CTMC) simulations are adequate to describe the processes. More specifically, CTMC calculations \citep{pohl} demonstrated that while previous rate coefficients obtained by Mansbach and Keck \citep{mansbach} 
are correct for large energy transfers, significant corrections, singular in $1/\Delta E$, have to be introduced 
for the proper description of collisions at small energy transfer. 

For collisional excitation, the proposed rate formula is \citep{pohl}

\begin{equation}
k_{if} = k_0 \epsilon_f^{3/2} \left[
\frac{22}{(\epsilon_i + 0.9)^{7/3}} + \frac{9/2}{\epsilon_i^{5/2} \Delta\epsilon^{4/3}}
\right] e^{\epsilon_f - \epsilon_i},
\label{Eq1}
\end{equation}
where $k_0 =\beta e^4/\sqrt{m {\cal R}}$ ($k_0$ expressed in ${\mathrm{cm}}^3/s$ in the cgs system with the electric charge in Gaussian units), $\epsilon_i = \beta E_i$, $\epsilon_f = \beta E_f$, 
with $E_i={\cal R}/{n_i}^2$, $E_f={\cal R}/{n_f}^2$ the absolute values of the initial and final energies, 
${\cal R}$ the Rydberg constant, and $\Delta \epsilon=\beta(E_f-E_i)$.

Equation (\ref{Eq1}) does not describe correctly the $\beta \rightarrow 0$ limit, because it has a 
power-like $\beta^{s}$ behavior as opposed to the much slower $\log(\beta)$ dependence expected by the Born approximation. This suggests to incorporate the expected Born-like behavior in the classical formula (\ref{Eq1}) to adequately describe the collision rate coefficients over the whole range of temperatures. By replacing the exponential factor $\exp(\epsilon_f - \epsilon_i)$ in (\ref{Eq1}) with the ``quantum factor'' $\Delta\epsilon\; \Gamma(0,\Delta\epsilon)$, where we have introduced the incomplete gamma function as

\begin{equation}
\Gamma(0, \Delta \epsilon)=\int_{\Delta \epsilon}^{+\infty} \frac{e^{-x}}{x} dx,
\end{equation}
we obtain an expression for the rate coefficient that has the correct behavior at both low and high temperatures, and maintains its validity even at large $n$. Moreover, the formula may be extended to low $n$ by introducing a simple fitting factor that is in the range of unity uniformly across all parameters, and that can be found by direct comparison with the accurate R-matrix results of Pryzbilla and Butler, for transitions between low  $n$ \citep{przybilla}. 

The resulting expression is \citep{VOS2}

\begin{equation}
k_{if} =  k_0 \left(\frac{\epsilon_f}{\epsilon_i}\right)^{3/2}  
\left[\frac{22}{(\epsilon_i + 0.9)^{7/3}} + \frac{9/2}{\epsilon_i^{5/2} \Delta\epsilon^{4/3}}\right] 
\left(\frac{3.5 + 0.18 n_f^2}{1 + 1/\epsilon_i^{5/2}}\right)\Delta\epsilon \;\Gamma(0,\Delta\epsilon).
\label{P2}
\end{equation}

\begin{figure}
\centering
\includegraphics[width=2.6in]{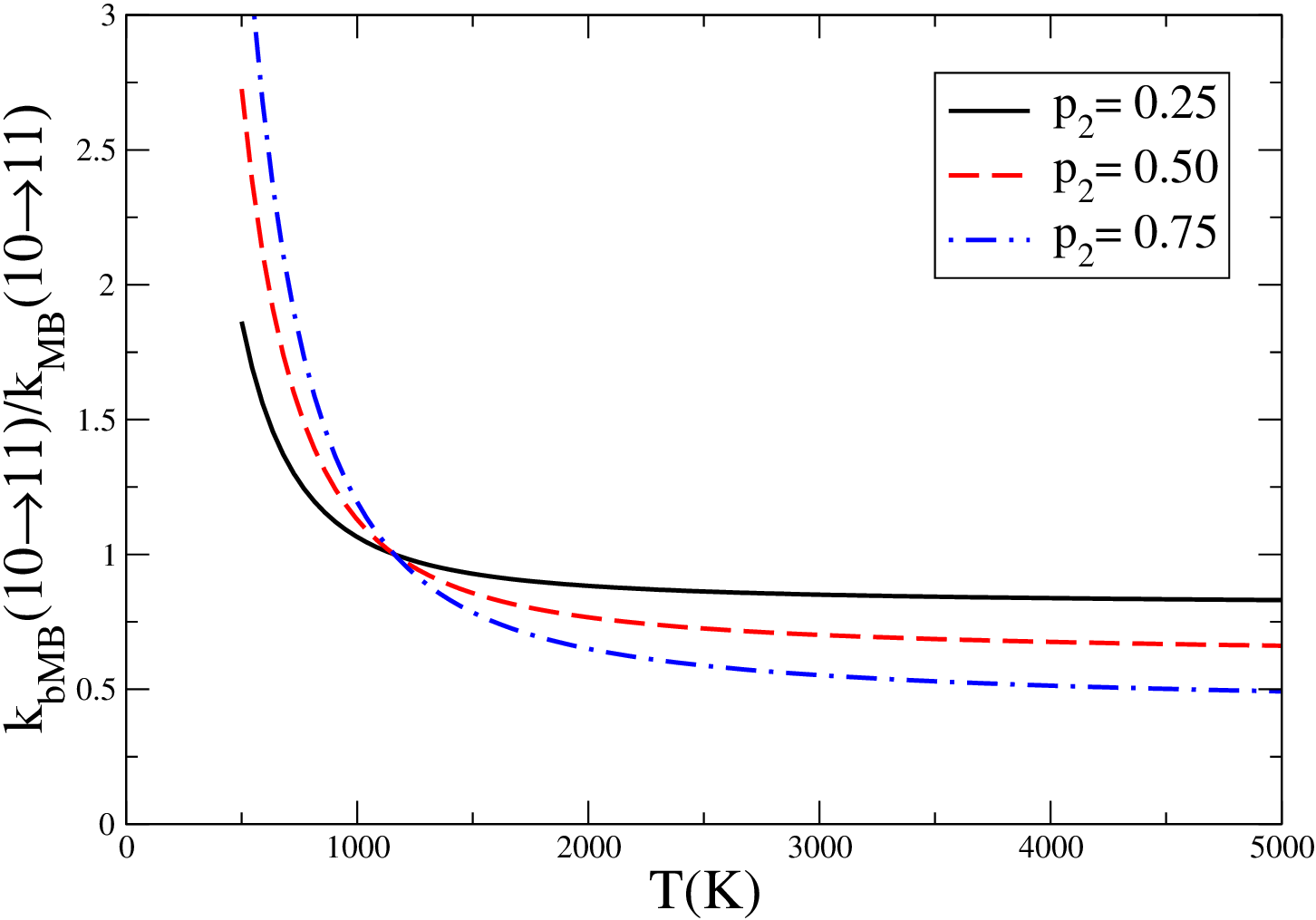}
\includegraphics[width=2.6in]{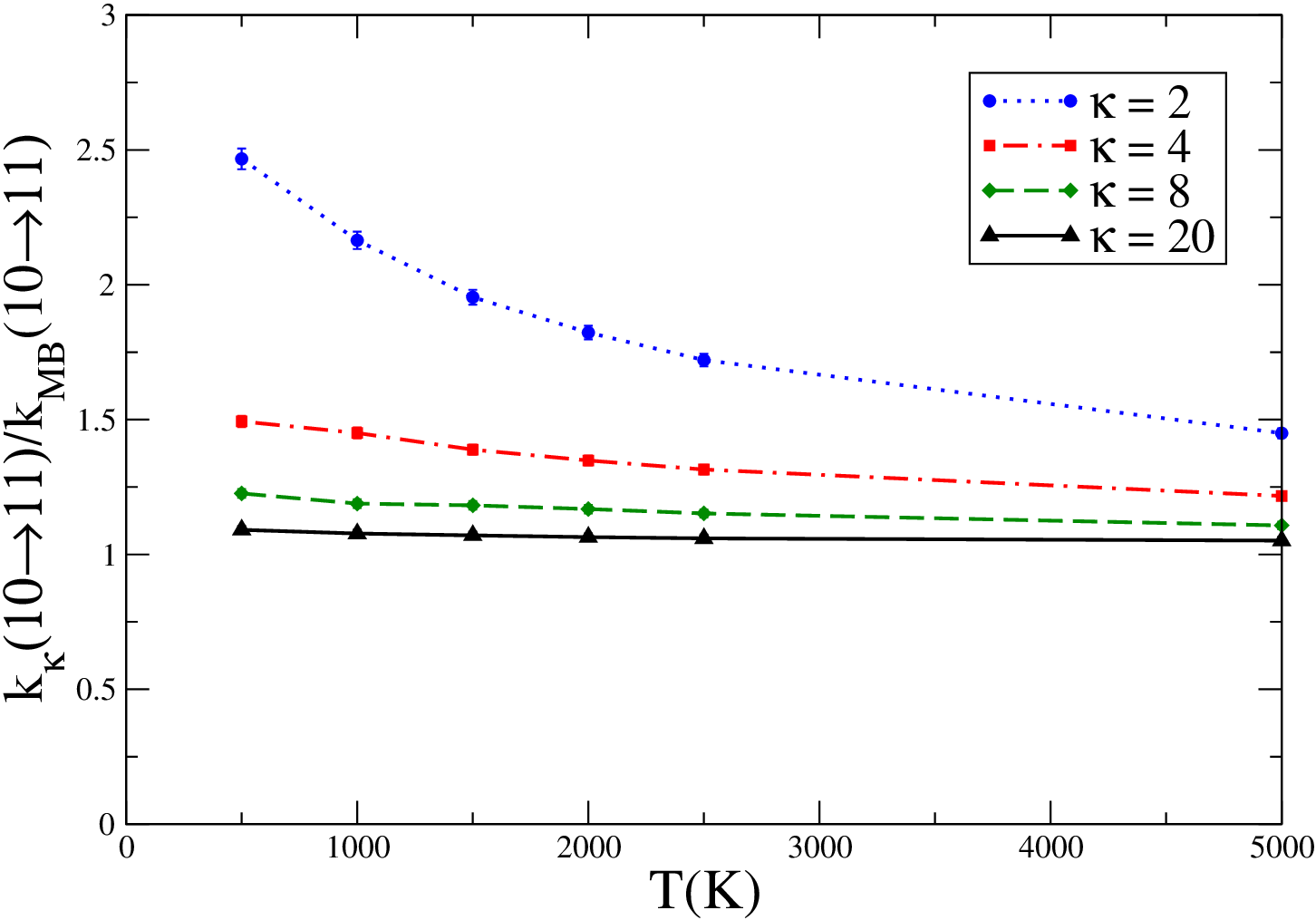}
\caption{\label{CTMC} Rate coefficients for H($n=10\rightarrow n'=11)$ Rydberg excitation as a function of temperature, in the 500-5000 K range, for bMB distribution with different weights for the two components and $T_2=10 T_1$ (left) and for $\kappa$ distributions with various $\kappa$ parameters and $\eta=0$ (right). Rate coefficients are scaled by the corresponding MB rates, obtained for $p_2=0$ for the bMB distribution, and $\kappa \rightarrow \infty$ for the $\kappa$-distribution. Notice the suppression of the rate coefficient for bMB occurring at high temperature with progressively increasing high-temperature ($T_2$) component, as the high-energy tail of this component becomes ineffective for the transition due to the $1/E$ dependence of the collision cross-section.}
\label{Fig2}
\end{figure}

In \citet{VOS2}, a comparison between this interpolating formula and many analytical models valid at low $n$ is carried out for various transitions and in a temperature range 
$2,500 < T < 250,000$ K. Although both plots and tables in \citet{VOS2} show disagreement between various models even by a factor two in some cases, it is still worth to explore the effect of bMB and $\kappa$-distributions within the same model especially considering that the cross section is quite sensitive to the details of the high-energy population. 

The evaluation of the rate coefficients for the bMB distribution is easily expressed as $k_{if}^{(bMB)} =  p_1 k_{if}^{(1)} + p_2 k_{if}^{(2)}$, denoting with $k_{if}^{(1)}$ and $k_{if}^{(2)}$ the rate coefficients for MB probability distributions at $\beta_1$ and $\beta_2$.
In figure \ref{Fig2} (left), we plot the rate coefficients for electron-hydrogen scattering versus temperature for the bMB distribution with different weights, all normalized to the analogous rate coefficients for a single MB distribution without the high-temperature component ($p_2=0$). 

The evaluation of the rate coefficients for the $\kappa$-distribution is considerably more involved, requiring extensive CTMC simulations similar to the ones carried out in \citet{pohl}, but starting from $\kappa$-distributed configurations. The initial positions and velocities for the Rydberg electrons are generated according to a microcanonical distribution corresponding to the H($n=10$) energy level, and a continuous distribution of angular momenta. The initial conditions for the incoming electrons are sampled with random impact parameters and velocities from a $\kappa$-distribution in (\ref{kappa}) with $\eta = 0$. The temperature can then be related to the most probable velocity, independent of $\kappa$, thereby allowing for a fair comparison between the different $\kappa$ distributions.
 
To obtain $\kappa$-distributed velocities, it is most straightforward to first sample a random number $u$ from a Fisher-Snedecor F-distribution \citep{Abramowitz}, $u \sim F(3, 2\kappa-1)$, and then assign the velocity as $v = v_T  \sqrt{3u(\kappa + \eta) /(2\kappa - 1) }$, with $v_T = \sqrt{2 k_B T/m}$. Most numerical packages have routines for dealing with the F-distribution  \citep{NumRep}. Alternatively, $\kappa$-distributed random numbers can be obtained by inverting the cumulative distribution function, as detailed in Appendix A. 

In figure \ref{Fig2} (right), results of the CTMC simulations are also shown, each simulation consisting of $4 \times 10^5$ complete trajectories, with random initial conditions generated according with the rules explained above, classified according to the observed outcome, and repeated for each combination of parameters $\kappa$ and temperature $T$. 
As $\kappa$ increases, the probability distribution of velocities for the electrons approaches a MB distribution and the rate coefficients scaled by the corresponding MB values, approach unity.

The most pronounced deviation is observed at low temperatures, when the rate coefficients can be as much as 2.5 times larger than the MB ones. In this specific example, the $10 \rightarrow 11$ transition for hydrogen corresponds to 28.3 meV, {\it i.e.} about 280 K. 
In order to remain in a scattering state, the impinging electron producing the transition should have an energy much larger than the excited electron. This implies that MB-distributed electrons in the low temperature range around 500 K are disadvantaged with respect to $\kappa$-distributed electrons with a larger high-energy population. As the temperature is increased, this disadvantage is progressively compensated. In the case of a bMB distribution as in Fig. 2 (left), the use of a high temperature component progressively results in even lower rate coefficients with respect to a single MB distribution as the temperature is increased. This result is reasonable considering that the collision cross sections decrease as $1/E$, and therefore too large energies for the impinging electrons are ineffective in causing the atomic transitions.

\section{Electron-Rydberg atom ionization rate coefficients}

We now analyze the case of ionization of the Rydberg atoms,  employing the generalized ionization cross sections for Rydberg atoms introduced in \citet{rost}
\begin{equation}
\sigma(E) = (1 + 1/\alpha)^{\alpha+1} \frac{E_M E^\alpha}{(E + E_M/\alpha)^{\alpha + 1}} \sigma_M,
\label{rostcross}
\end{equation}
where $E$ is the excess energy, {\it i.e.} the difference between the absolute energy and the ionization threshold energy, and $\alpha$ is the Wannier threshold exponent \citep{wannier} characteristic of each target-projectile system at low energies ($\alpha=1.127$ for the electron-hydrogen collision). This parameterization of the cross section has the proper low energy Wannier threshold behavior, $\sigma(E\rightarrow 0) \sim E^\alpha$, the expected behavior $\sigma \sim 1/E$ for $E\rightarrow \infty$, and peaks for $E = E_M$ such that $\sigma(E_M) = \sigma_M$.
Notice that the $1/E$ asymptotic behavior is reminiscent of a classical approximation, and therefore does not account for the logarithmic quantum corrections, as would be expected in the Born approximation. 

\begin{figure}
\centering
\includegraphics[width=2.6in]{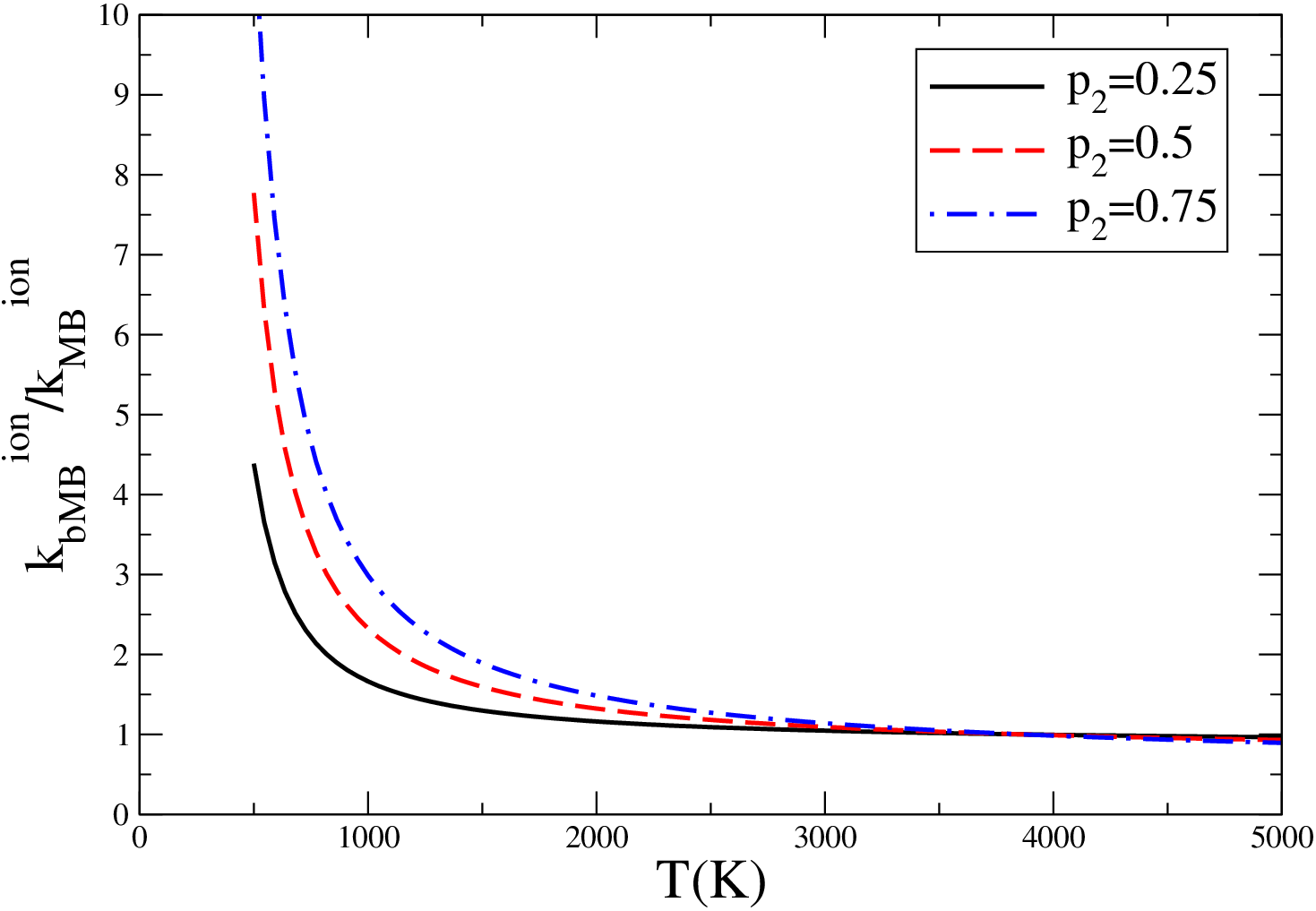}
\includegraphics[width=2.6in]{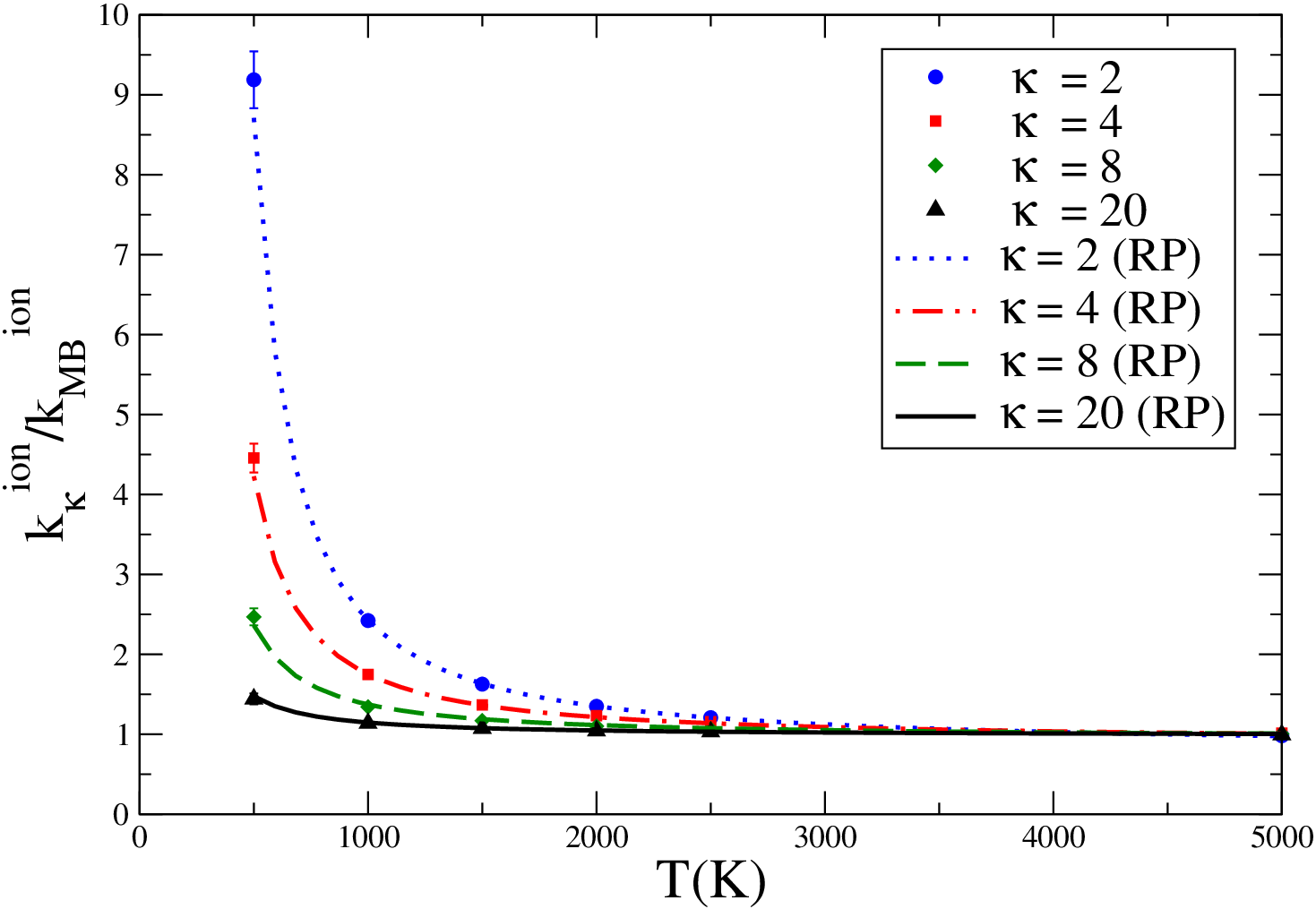}
\caption{Ionization rate coefficients, normalized to the corresponding MB rates, for various  bMB distributions with $T_2=10~ T_1$ (left), and for various $\kappa$-distributions (right). The dots in the right plot originate from CTMC simulations, including their statistical errors, while the lines are from the numerical integration of (\ref{rates}) with the model ionization cross section of (\ref{rostcross}) taken from the Rost and Pattard (RP) parameterization. The parameters used are: $n=10$, $\eta = 0$, $\alpha = 1.127$, and $E_M = 2 I_n$, where $I_n$ is the binding energy of the Rydberg atom in state with quantum number $n$. The RP rate coefficients for the $\kappa$ distributions strongly depend on the $E_M/I_n$ ratio since a resonance occurring at higher energies has better overlap with the enhanced high-energy tail of the $\kappa$ distribution. For instance, by choosing $E_M=5 I_n$ we obtain RP rate coefficients larger by a factor 1.8 with respect to the $E_M=2 I_n$ case for the $\kappa=2$ distribution.}
\label{Fig3}
\end{figure}

In figure \ref{Fig3} (left) we plot the ratio between the rate coefficient for a mixture of bMB distribution with different values for $p_2$, and the corresponding rate coefficient  for $p_2=0$ versus temperature. The calculations were  repeated for $\kappa$-distributions with $\eta=0$ and the outcome is depicted in figure (\ref{Fig3}) (right). In addition to the analysis based on \citet{rost}, we have also evaluated the rate coefficients based on the same CTMC simulations performed for the excitation analysis. We have chosen an energy for the peak of the cross section $E_M=2 I_n$ as suggested by numerical simulations \citep{vrinceanu}. This is also in line with various parameterization of experimental data showing that the peak of the ionization cross-section is of the order of 5-10 times the ionization energy for low $n$, thereafter approaching a value of 2-3 times the ionization energy at higher $n$ quantum numbers \citep{janev1}.

The agreement between the simulations and the analytical interpolation based on \citet{rost} is remarkably good, and yet not completely surprising because they both 
share the classical behavior at high energy, omitting Born-like logarithmic corrections to the cross-section. There are large (up to an order of magnitude at the lowest $\kappa$) deviations at low temperature, in a range of interest for plasma diagnostics in the scrape-off layer region and in the divertor region of a tokamak.

The effect of a more pronounced high-energy tail becomes minimal at high temperature, as expected for phenomena in which a cross section with a resonant behavior appears. However, it is worth to remark that we expect suppression of the rate coefficient in some region of the parameter space for both bMB and $\kappa$ distributions. This behavior is emphasized by a side to side comparison for two $\kappa$ distributions, corresponding to different $\eta$ parameters, in figure 4. Notice that for $\eta=1/2$, a slight suppression occurs, in the discussed temperature range, only for the $\kappa=2$ case and at the highest temperature. In the case of $\eta=-3/2$ instead the suppression is visible for all  values of $\kappa$, reaching almost a factor of two for $\kappa=2$ at about 2,000 K. The suppression is not monotonic, an effect already observed in figure 5 in \citet{nicholls1}.
This effect is easy to explain in terms of mismatching between the resonance of the cross section of the process and the energy probability distribution. This is more evident in the case of $\eta=-3/2$ and small $\kappa$, as the energy probability distribution has a large excess at both low and high energies, as visible in panels (c) and (f) of figure \ref{Fig1}, {\it i.e.} far from the resonance condition for the cross-section. 
The same suppression is also present in bMB distributions, as barely evident in figure 3 a. In appendix B, we discuss an analytical model based on a convenient parameterization of the resonant cross-section, leading to a simple formula showing enhancement and suppression of the corresponding rate coefficient in the opposite temperature limits. 

\begin{figure}
\centering
\includegraphics[width=2.6in]{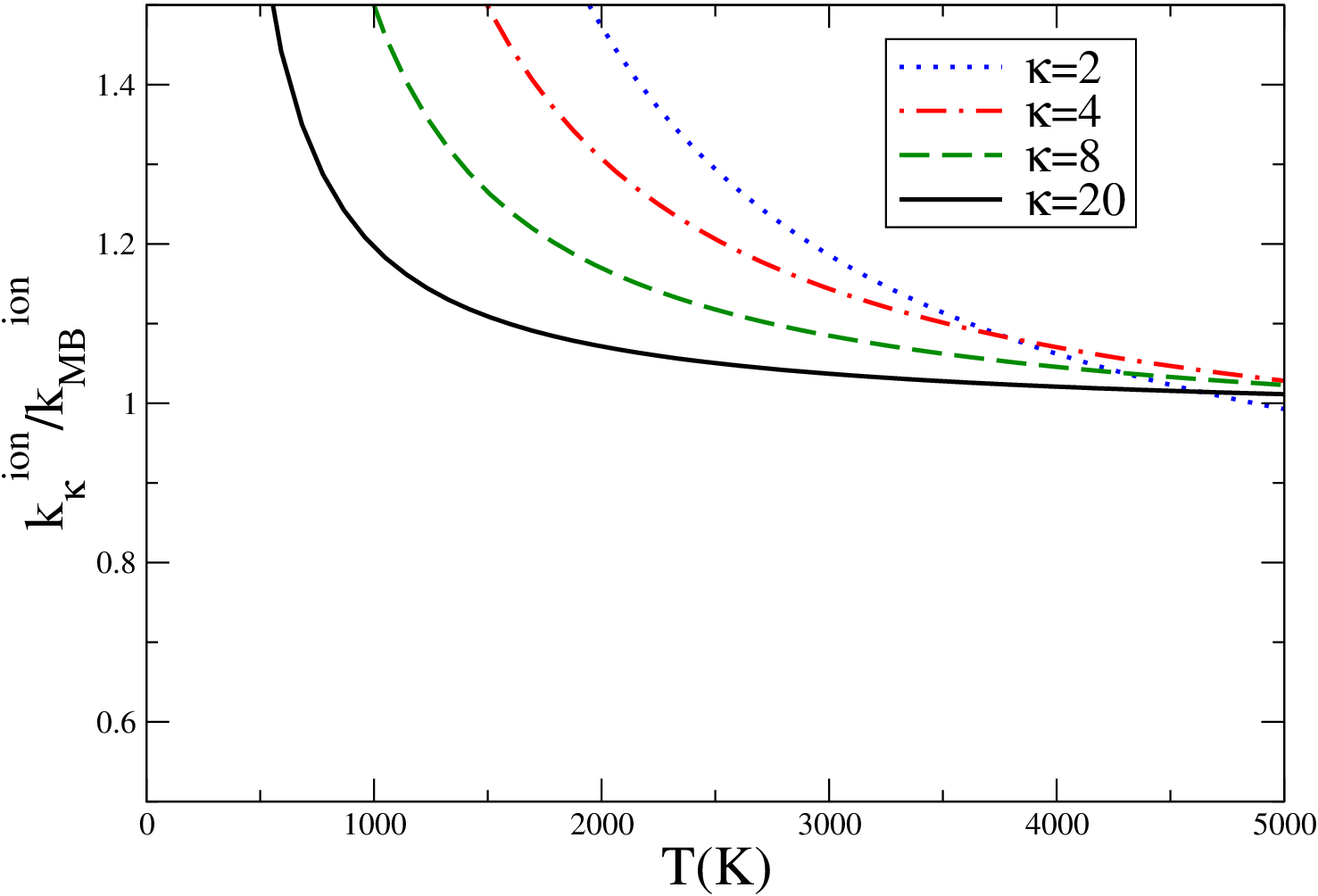}
\includegraphics[width=2.6in]{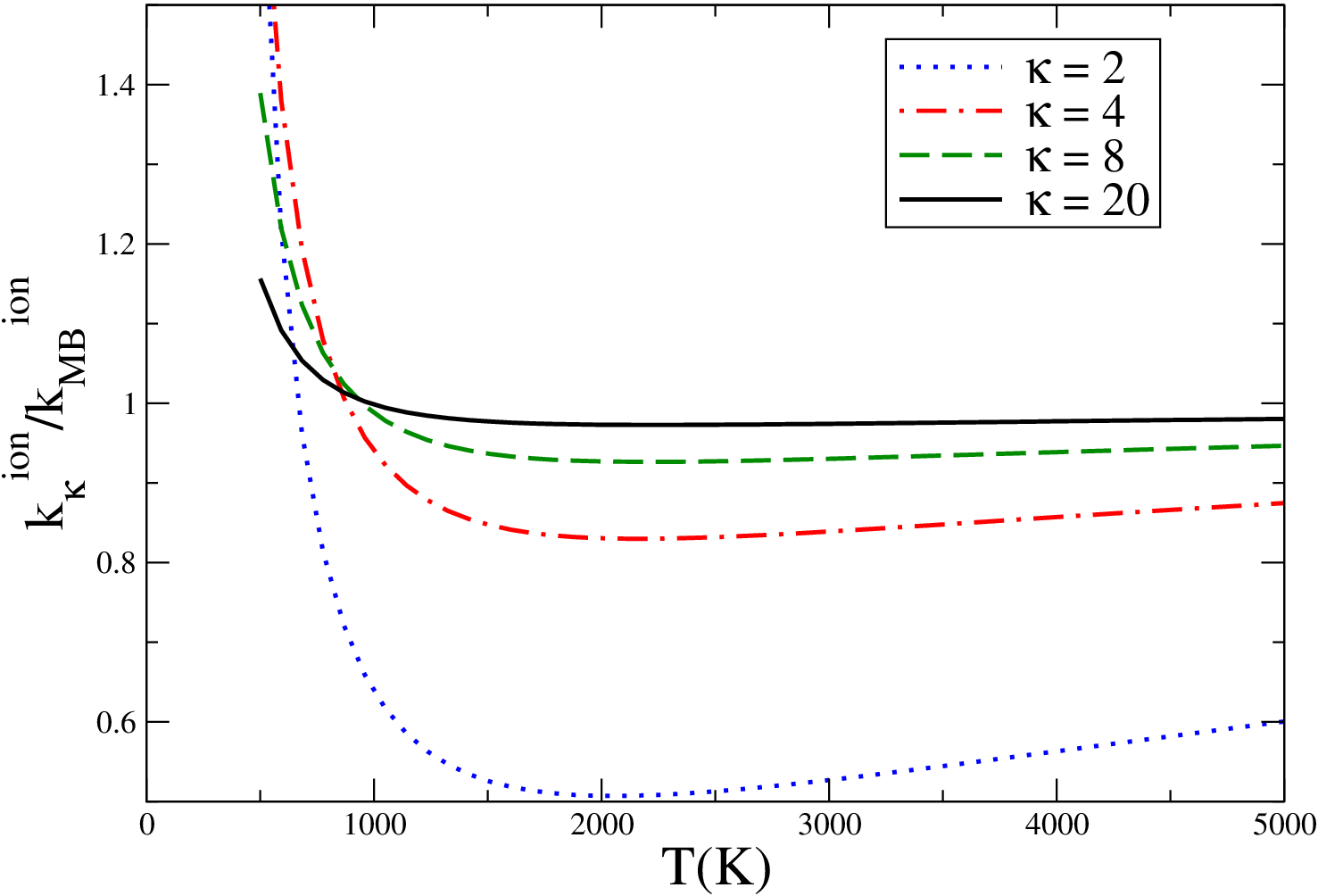}
\caption{Evidence for the suppression of the ionization rate coefficients with the $\kappa$ distributions. The rate coefficients are normalized to the corresponding MB rates, for $\eta=1/2$  (left) and $\eta=-3/2$ (right), for various values of $\kappa$.}
\label{Fig4}
\end{figure}

\section{Conclusions}

We have discussed the sensitivity of the rate coefficients for excitation and ionization in electron-atom
collisions to deviations from Maxwell-Boltzmann distributions. The outcome indicates that the rate coefficients 
may differ significantly from those derived from MB distributions especially at low temperatures, in the range of 1,000 K. 
The results readily translate into rate coefficients for proton-atom collisions, since there is an approximate scaling with the square mass ratio in the 
excitation case, and analogous parameterization of the cross-section with the Rost and Pattard approach, using  different Wannier exponents, for the ionization case. 

Our discussion is of interest to tokamak physics for two main reasons. First, an accurate knowledge of 
the kinetics of the involved ions and atoms in the plasma-edge region is a crucial element for the correct working of ITER and related 
fusion reactors, and for this reason a dedicated facility, DTT, is under construction in Frascati \citep{albanese} with the aim to guide 
the divertor design for the ITER facility. In the plasma edge region the plasma is colder than in the core of the reactor, and contains 
impurities due to the interactions with the materials of the vessel-wall. Detailed models of this region are important because these 
boundary effects regulate the amount of impurities penetrating the core, determining the plasma heat load on the divertor target plates, and therefore the overall performance of the reactor \citep{reiter,winter}.
Second,  the  extraction of the temperature and fluxes for plasma components can be affected by this systematic effect, since so far all calculations 
for the flux of charge-exchanged fast neutrals escaping from a plasma assume MB distributions \citep{tugarinov,hollmann}. 

We believe that the 
$\kappa$ distributions may play a major role in the atomic physics around the plasma-edge region, since the conditions for local 
thermal equilibrium are not easily met due to the steep drop in plasma density.  
While our study is limited to two specific atomic processes, a more general programme should also be implemented for charge exchange and recombination processes \citep{takamura}. 

\acknowledgments
DV is grateful to Texas Southern University High Performance Computing Center for making the necessary 
computational resources available, and to the National Science Foundation for the support received 
through grants PHY-1831977 and HRD-1829184. 
This work was also partially supported by the National Science Foundation through a grant for the Institute 
for Theoretical Atomic, Molecular and Optical Physics at Harvard University and the Smithsonian 
Astrophysical Laboratory.

\appendix

\section{Cumulative functions for generating $\kappa$-distributed velocities}

Monte Carlo Classical Trajectory simulations require $\kappa$-distributed velocities corresponding to the energy probability density (\ref{kappa}), and we summarize here some of their properties and how to generate them from uniform random numbers.
The normalized velocity probability density in three dimensions is defined by
\begin{equation}\label{kappa-speed}
f_{3D}(v; \kappa, \eta) = \frac{4}{\sqrt{\pi}} \left(\frac{\beta m/2}{\kappa +\eta}\right)^{3/2} \frac{\Gamma(\kappa +1)}{\Gamma(\kappa - 1/2)} \frac{v^2}{\left(1 + \frac{\beta m v^2/2}{\kappa + \eta}\right)^{\kappa+1}},
\end{equation}
where $\beta$ is the inverse temperature and $m$ is the particle mass. The parameter $\eta$ allows for  selecting different classes of $\kappa$-distributions that have special properties and interpretations, as explained below. The conventional MB distribution is obtained from (\ref{kappa-speed}) in the $\kappa\rightarrow \infty$ limit.

The $\kappa$-distribution in (\ref{kappa-speed}) peaks at a velocity given by $v_p = \sqrt{2 (\kappa + \eta)/(\kappa \beta m)}$. This shows that the class of $\kappa$-distributions with $\eta = 0$ have the maximum at a $\kappa$-independent velocity. Therefore for this class of distributions the temperature can be defined in a $\kappa$-independent way as $T = m v_p^2/(2 k_B)$, with $v_p$ the most probable velocity.

The mean and mean square velocities for $\kappa$-distributions respectively are
\begin{equation}
\langle v \rangle = 2 \sqrt\frac{2(\kappa + \eta)}{\pi \beta m} \frac{\Gamma(\kappa - 1)}{\Gamma(\kappa - 1/2)}\quad\mbox{and}\quad \langle v^2 \rangle = \frac{3(\kappa + \eta)}{2\kappa - 3} \frac 2{\beta m}\;.
\end{equation}
For the class of distributions  with $\eta = -3/2$ the temperature can be defined in a $\kappa$-independent way as $T = m \langle v^2 \rangle / (3 k_B)$.
For a general $\kappa$-distribution, the ``core'' temperature, obtained from the most probable velocity, and the ``kinetic'' temperature that results from the mean square velocity are different, and depend on the specific $\kappa$ and $\eta$ parameters. Only for the special values $\eta = 0$, and $\eta = -3/2$, the interpretation of temperature becomes straightforward.
Another important class of distributions is $\eta = 1/2$ where the most probable energy in the distribution (\ref{kappa}) relates with the temperature regardless of $\kappa$, essentially the same as in the MB distribution obtained for $\kappa \rightarrow \infty$.

The cumulative distribution function related to (\ref{kappa-speed}) can be calculated exactly as
\begin{eqnarray}
F(x; \kappa, \eta) &=& \int_0^x   f_{3D}(t;\kappa, \eta) \; dt = \frac{1}{x} \sqrt{\frac{\kappa + \eta}{\pi}} \frac{\Gamma(\kappa+1)}{\Gamma(\kappa+3/2)} 
\frac{1}{(1+x^2/(\kappa + \eta))^{\kappa}} \times \nonumber\\
& & \left[ {}_2F_1\left(1, -\kappa -\frac{1}{2}, \frac{1}{2}; -\frac{x^2}{\kappa + \eta}\right) - 1 - \frac{2\kappa+1}{\kappa + \eta} x^2\right], \label{cdf}
\end{eqnarray}
where $x^2 = \beta m v^2/2$, and ${}_2F_1$ is the ordinary hypergeometric function.
Starting from uniformly distributed random numbers $u$ and using the inverse of the cumulative distribution function (\ref{cdf}) one obtains $\kappa$-distributed velocities using $v = F^{-1}(u)\sqrt{2/(\beta m)}$.
A practical way to solve the transcendental equation $F(x)= u$ is to create a two-way table for $x$ and $u$, and then use an interpolation to obtain $x$ for any $0 < u < 1$.

For one-dimensional problems a $\kappa$-distribution of velocities has a probability density function
\begin{equation}
f_{1D}(v; \kappa, \eta)=\sqrt{\frac{\beta m}{2\pi (\kappa + \eta)}} \frac{\Gamma(\kappa+1)}{\Gamma(\kappa+1/2)}
\frac{1}{\left(1+ \frac{\beta m v^2/2}{\kappa+\eta}\right)^{\kappa+1}}.
\end{equation}
This distribution has a maximum at $v=0$, with zero average $\langle v \rangle=0$ and variance 
$\sigma^2=\langle v^2 \rangle = (\kappa + \eta)/(\beta m(\kappa-1/2))$.  The special class of $\kappa$ distributions with $\eta = -1/2$ has a $\kappa$-independent variance, which  leads to a direct interpretation of temperature as $T = m\langle v^2\rangle/k_B$. As expected, for large $\kappa$ the 
distribution becomes normal, $\lim_{\kappa \rightarrow \infty} f_{1D}(v; \kappa, \eta)=\sqrt{\beta m/(2 \pi)} \exp(-\beta m v^2/2)$. 
The corresponding cumulative distribution function can be expressed as
\begin{eqnarray}
F_{1D}(x; \kappa, \eta) &=& \int_{-\infty}^x f_{1D}(t; \kappa, \eta) dt = \nonumber \\ 
& & \frac{1}{2} + \frac x{\sqrt{(\kappa + \eta)\pi}} \frac{\Gamma(\kappa +1)}{\Gamma(\kappa+1/2)}  {}_2F_1\left(\frac{1}{2}, \kappa+1, \frac{3}{2}, -\frac{x^2}{\kappa + \eta} \right),
\end{eqnarray}
with $x^2 = \beta m v^2/2$. An alternative method to generate 1D $\kappa$-distributed velocities uses random variates of the Student t-distribution, as described in \citet{Abdul2014}.
This distribution is not useful to generate configurations in three dimensions because the magnitude of the sum of three squares of 1D $\kappa$-distributed velocities $\sqrt{v_1^2+v_2^2+v_3^2}$ is not $\kappa$-distributed. This does not happen for MB distributions since  a three-dimensional MB distribution is factorizable into three one-dimensional normal distributions. 

\section{Enhancement and suppression of scattering rates in non-MB distributions: An analytical example}\label{appC}

As visible in the left plot of figure 2 and, to a smaller extent, in the analogous one in  figure 3, the rate coefficients for bMB distributions can be enhanced or suppressed depending on the temperature range. Intuitively this depends on the overlap between the energy distribution and the cross-section dependence on energy. In particular, for resonant cross-sections, one expects that at high temperature the overlap in the presence of enhanced high-energy tails is smaller with respect to the corresponding MB distribution thereby suppressing the rate coefficient. In order to show this analytically we discuss the case of a simple cross-section parameterized as

\begin{equation}
\sigma(E) = \sigma_M \left(\frac{E}{E_M}\right)^\lambda \exp[-\beta_\lambda (E-E_M)],
\label{Crossection}
\end{equation}
where $E_M=\lambda/\beta_\lambda$ is the value of the energy for which the cross-section peaks, and $\beta_\lambda, \lambda>0$. 
While this cross-section is not encountered in concrete physical applications, it has the same 
features of many realistic cross-sections, with a peak value at some intermediate energy and tails falling 
at lower and higher energies. Its advantage is that it allows to get a simple 
expression for the rate coefficient when convoluted with a MB energy distribution
at inverse temperature $\beta_1$
\begin{equation}
P_{MB}(E) = 2\sqrt{\frac{E}{\pi}}\beta_1^{3/2} \exp(-\beta_1 E),
\label{MB}
\end{equation}
obtaining a rate coefficient, based on (\ref{rates})
\begin{equation}
k_1 = \sqrt{\frac{8}{\pi m}} \frac{\sigma_M \exp{(\beta_\lambda E_M)}}{E_M^\lambda} \beta_1^{3/2} 
(\beta_1+\beta_\lambda)^{-\lambda-2} \Gamma(\lambda+2).
\label{RateMB}
\end{equation}

For a bMB mixture involving another inverse temperature $\beta_2$ and 
weights $p_1$ and $p_2$ we then obtain, by defining the common factor 
$C=\sqrt{8/(\pi m)} \sigma_M \exp(\beta_\lambda E_M)/E^\lambda_M  \Gamma(\lambda+2)$

\begin{equation}
k_{bMB} = C \left[
p_1 \beta_1^{3/2} (\beta_1+\beta_\lambda)^{-\lambda-2}+
p_2 \beta_2^{3/2} (\beta_2+\beta_\lambda)^{-\lambda-2}\right].
\label{RatebMB}
\end{equation}
By taking the ratio between (\ref{RatebMB}) and (\ref{RateMB}), with 
$\beta_2/\beta_1=\xi={\mathrm{const}}<1$ the rate coefficient ratio is written as

\begin{equation}
\frac{k_{bMB}}{k_{MB}}=p_1+(1-p_1) \xi^{3/2} 
\left(\frac{\xi \beta_1+\beta_\lambda}
{\beta_1+\beta_\lambda}\right)^{-\lambda-2}.
\end{equation}
In the (low-temperature) limit $\beta_1 \rightarrow \infty$ we have  
$k_{bMB}/k_{MB} \rightarrow p_1 +(1-p_1) \xi^{-\lambda-1/2} > 1$. 
In the opposite limit, $\beta_1 \rightarrow 0$, the rate coefficient ratio instead tends to 
$k_{bMB}/k_{MB} \rightarrow p_1 +(1-p_1) \xi^{3/2} < 1$.

\end{document}